\documentclass[aps,
twocolumn, 
showpacs,
prx,
superscriptaddress,
tightenlines,
10pt,
longbibliography] 
{revtex4-2}

\usepackage{graphicx}
\usepackage{color,soul}
\usepackage{xcolor}
\usepackage{braket}
\usepackage{hyperref}
\usepackage{comment}
\usepackage{amsmath}
\usepackage{appendix}
\usepackage{physics}
\usepackage[separate-uncertainty=true]{siunitx}
\usepackage{microtype}
\usepackage{import}
\usepackage{xifthen}
\usepackage{pdfpages}
\usepackage{transparent}
\graphicspath{{figures/}}
\usepackage{mwe}

\def\g2{g^{(2)}(0)}

\makeatletter
\newcommand*{\defeq}{\mathrel{\rlap{%
                     \raisebox{0.3ex}{$\m@th\cdot$}}%
                     \raisebox{-0.3ex}{$\m@th\cdot$}}%
                     =}
\makeatother

\makeatletter
    \renewcommand\@make@capt@title[2]{%
     \@ifx@empty\float@link{\@firstofone}{\expandafter\href\expandafter{\float@link}}%
      {\textbf{#1}}\@caption@fignum@sep#2\quad}%
    \makeatother

\UseRawInputEncoding
\usepackage{lipsum}
\makeatletter
\AtBeginDocument{\let\LS@rot\@undefined}
\makeatother

\usepackage[T1]{fontenc}
\usepackage{libertinus}
\usepackage{libertinust1math}
\usepackage[cal=stixtwoplain, bb=pazo]{mathalpha}

\usepackage{microtype}
\usepackage{mathtools}
\usepackage{bm}

\usepackage{xcolor}
\usepackage{graphicx}
\graphicspath{{./img/}}

\usepackage{enumitem}
\usepackage{booktabs}

\let\xtimes\times
\let\times\cdot
\usepackage{csquotes}
\usepackage[
  input-comparators={<=>\approx\ge\geq\gg\le\leq\ll\sim\lesssim},
  range-phrase={\,\text{--}\,},
  product-symbol=\xtimes,
  ]{siunitx}
\usepackage[version=4]{mhchem}

\usepackage{hyperref}
\usepackage[all]{hypcap}
\usepackage[ocgcolorlinks]{ocgx2}

\makeatletter
  \patchcmd\frontmatter@setup{\normalfont}{\normalfont\sffamily}{}{}
  \patchcmd\frontmatter@title@format{\bfseries}{\sffamily\bfseries}{}{}
  \patchcmd\frontmatter@affiliationfont{\it}{\slshape}
  \patchcmd\frontmatter@title@format{\large}{\Large}{}{}
  \patchcmd\section{\bfseries}{\sffamily\bfseries}{}{}
  \patchcmd\subsection{\bfseries}{\sffamily\bfseries}{}{}
  \patchcmd\subsubsection{\itshape}{\sffamily}{}{}
  \patchcmd\@makecaption{\rmfamily}{\sffamily}{}{}
\makeatother

\makeatletter
  
  \pretocmd\@make@capt@title{\begingroup\bfseries}{}{}
  \patchcmd\@make@capt@title{\@caption@fignum@sep}{\@caption@fignum@sep\endgroup}{}{}
  \def\@caption@fignum@sep{: }

\makeatother

\makeatletter
  \newcommand*\@withperiod[1]{#1.}
  \patchcmd\paragraph{\normalfont\normalsize\itshape}{\normalfont\normalsize\itshape\@withperiod}{}{}
\makeatother

\makeatletter
  \def\bibsection{%
    \expandafter\section\expandafter*\expandafter{\refname}%
    \@nobreaktrue
    \raggedright
  }%
\makeatother

\definecolor{hyperintcolor}{rgb:HTML}{416b16}
\definecolor{hyperextcolor}{rgb:HTML}{2263a3}
\hypersetup{
  colorlinks,
  linkcolor=hyperintcolor,
  citecolor=hyperintcolor,
  filecolor=hyperextcolor,
  urlcolor=hyperextcolor,
}

\RenewDocumentCommand\doi{ m }{%
  \href{\doibase #1}{\nolinkurl{#1}}%
}

\makeatletter
  \def\mathopsym#1{\mathop{\kern\z@#1}}

  \def\@difsym{{\mathrm d}}
  \def\@delsym{\partial}
  \NewDocumentCommand\dif{s s o m}{%
    \IfBooleanF{#2}{\IfBooleanTF{#1}{\unskip}{\mathop{}}\!}%
    \@difsym%
    \IfValueT{#3}{^{#3}}%
    #4%
    \IfBooleanT{#1}{\mathop{}\!}%
  }
  \NewDocumentCommand\del{o m}{%
    \@delsym_{\mkern-2.5mu\relax#2}%
    \IfValueT{#1}{^{#1}}%
  }
  \NewDocumentCommand\evalwith{O{\big} m}{%
    \mathop{}\!#1|_{#2}%
  }
\makeatother

\DeclareSIUnit\sqr{\ensuremath{\square}}

\begin{document}
\title{Collective vacuum-Rabi splitting with \\ an atomic spin wave coupled to a cavity mode}

\author{F\'elix Hoffet}
\email{felix.hoffet@icfo.eu}
\affiliation{ICFO -- Institut de Ciencies Fotoniques, The Barcelona Institute of Science and Technology, Spain}

\author{Alexey Vylegzhanin}
\affiliation{Okinawa Institute of Science and Technology Graduate University, Onna, Okinawa 904-0495, Japan}

\author{Emanuele Distante}
\affiliation{ICFO -- Institut de Ciencies Fotoniques, The Barcelona Institute of Science and Technology, Spain}

\author{Lukas Heller}
\affiliation{ICFO -- Institut de Ciencies Fotoniques, The Barcelona Institute of Science and Technology, Spain}

\author{S\'{i}le Nic Chormaic}
\affiliation{Okinawa Institute of Science and Technology Graduate University, Onna, Okinawa 904-0495, Japan}

\author{Hugues de Riedmatten}
\affiliation{ICFO -- Institut de Ciencies Fotoniques, The Barcelona Institute of Science and Technology, Spain}
\affiliation{ICREA -- Instituci\'o Catalana de Recerca i Estudis Avan\c cats, 08015 Barcelona, Spain}

\begin{abstract}
A promising platform for quantum information research relies on cavity coupled atomic spin-waves, enabling efficient operations such as quantum memories, quantum light generation and entanglement distribution. 
In this work, we study the strong coupling between non-classical collective spin excitations generated by Raman scattering in a cold $^{87}\mathrm{Rb}$ atomic ensemble, and a single cavity mode. 
We report on an intracavity spin wave to single photon conversion efficiency of up to $\chi=0.75 \pm 0.02$ in the quantum domain, as evidenced by a violation of the Cauchy-Schwarz inequality.
Our work establishes a relationship between the retrieval of an atomic spin wave in the non-classical regime and the vacuum-Rabi splitting. We show that this relationship emerges within the efficiency spectrum, and we finally provide the optimal operational conditions to achieve the maximum intrinsic retrieval efficiency. 
Our data is well reproduced by simulations based on optical Bloch equations.
This work deepens the understanding of cavity-enhanced spin wave readout and its potential applications.
\end{abstract}

\maketitle

\section{Introduction}
Establishing efficient and controllable light-matter coupling is essential for quantum networks, where flying optical photons carrying quantum information are coupled into and out of matter-based quantum memories at network nodes \cite{Wehner2018}. 

One possibility to achieve strong light-matter coupling is to use single emitters, such as single ions or neutral atoms, positioned at the center of a high-finesse optical cavity. 
While this approach led to the demonstration of key quantum network primitives \cite{Reiserer2015} - including quantum repeater nodes \cite{Langenfeld2021, Krutyanskiy2023} and multiplexed memory registers \cite{Hartung2024, Krutyanskiy2024} - the weak free space atom-photon coupling strengths requires small-volume and high-finesse cavities, which introduces significant experimental challenges such as complicated in-vacuum cavity design, and restricted optical access for trapping and controlling the atoms or ions.

An alternative approach employs atomic ensembles, where the collective interaction between light and a large number of atoms enables strong light-matter coupling \cite{Sangouard2011}. In this case, photons are stored as collective atomic excitations, or spin-waves, which rely on the constructive interference among all the atoms to efficiently couple light in and out of a specific optical mode. 
This approach has demonstrated elementary instances of quantum repeaters \cite{Chou2005,Chou2007,Yuan2008}, efficient \cite{Laurat2006} and indistinguishable \cite{Felinto2006, Hoffet2024} single-photon generation , efficient quantum memories \cite{Wang2019, Cao2020} and multiplexed quantum nodes \cite{Tian2017a,Pu2017,Chang2019}. 
Since the interference scales with the number of atoms, achieving high storage efficiency requires either a large dilute ensemble or a small high-density one \cite{Gorshkov2007}. 
Extended atomic clouds require advanced atomic trapping techniques \cite{Hsiao2014} and suffer from spatially inhomogeneous stray fields across the ensemble that might limit memory coherence time, while high atomic density can also introduce limitations arising from atomic collisions \cite{Schmidt-Eberle2020}. 

\begin{figure*}[t!]
    \begin{minipage}[c]{0.9\linewidth}
   \def\svgwidth{\columnwidth}
   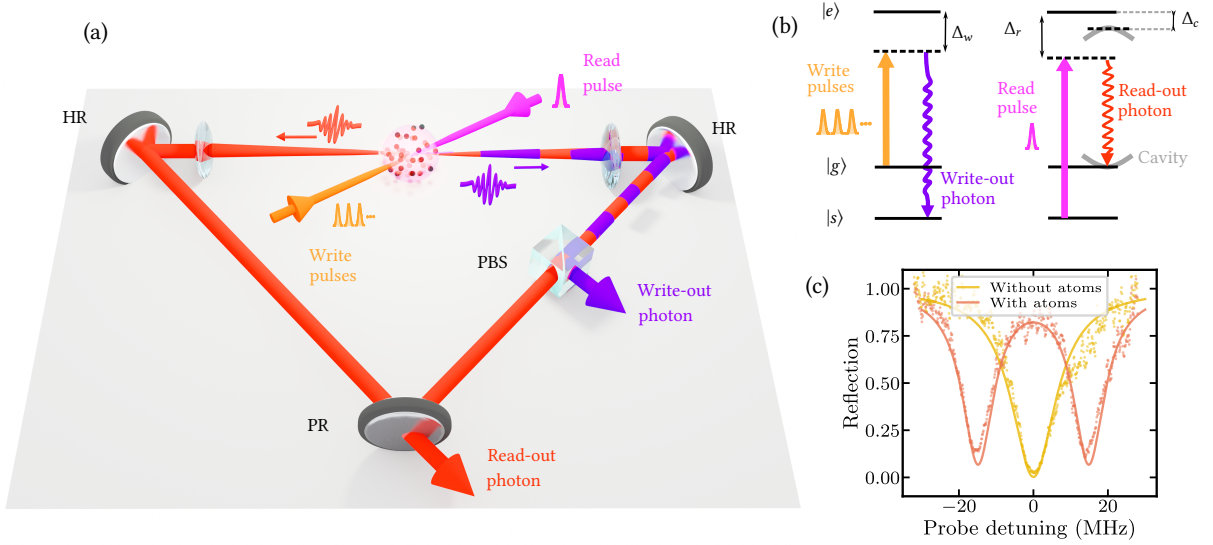
    \end{minipage}
    \caption{(a): Schematic view of the experiment. A ring cavity of finesse $\mathcal{F}=23$ is formed by two highly reflective mirrors (HR) and one partially reflective (PR) mirror ($86\%$ reflection).
    An atomic ensemble containing $N \sim 4\times10^5$ atoms is coupled to the cavity mode. Utilizing a flip-mirror mount (not shown), we can bypass the cavity to probe our cloud in single-pass if needed. The cavity is tuned to resonance with the read-out photon of the DLCZ protocol, while the write-out photon is out-coupled from the cavity using a polarization beam splitter (PBS). 
    (b): Relevant level scheme and laser frequencies for the DLCZ protocol. Here, ${\ket{g} = \ket{5S_{1/2}, F=2}}$, ${\ket{e} = \ket{5P_{3/2}, F=2}}$ and  ${\ket{s} = \ket{5S_{1/2}, F=1}}$. 
    (c): By measuring the reflection off the cavity of an input probe around the atomic resonance $\ket{g} \xrightarrow{} \ket{e}$, we observe the vacuum-Rabi splitting (red) of the bare cavity resonance (yellow). 
    The data is well reproduced by a semi-classical model (solid lines).
    }
    \label{fig:setup}
\end{figure*}

A promising solution combines the two strategies: using a low-finesse optical cavity to enhance the collective interaction between a moderately sized atomic ensemble and light. 
This hybrid approach significantly reduces experimental requirements, allowing the use of a simple magneto-optical trap and a large cavity design to be built outside the vacuum chamber, providing extensive optical access. 
This system has already been used to demonstrate key applications in quantum networking, such as the efficient generation of heralded single photons from spin waves \cite{Simon2007a, Bimbard2014a, Bao2012, Wang2023c, Cox2019}, temporally multiplexed atomic memories \cite{Heller2020}, the generation of light-matter entanglement \cite{Wang2021, Yu2020c}, and the development of field-deployable quantum network technologies \cite{Liu2024}. 
In combination with highly excited Rydberg states, recent experiments have also shown the capability for highly efficient photon-photon quantum gates \cite{Stolz2022a, Vaneecloo2022a}.

Here, we use this hybrid strategy to study the coupling of an atomic collective spin excitation to a single photonic cavity mode. 
Albeit with a moderate cavity finesse, our system reaches the strong-coupling regime owing to the collective atom-cavity interaction that surpasses both the cavity and the atom decay rates, as demonstrated by the reflectance spectrum of the coupled system. 
We use this strong interaction to enhance the retrieval efficiency of a DLCZ quantum memory \cite{Duan2001}. 
Here, a spin wave is first created in a cold atomic $^{87}\mathrm{Rb}$ ensemble,  heralded by a photon detection. A subsequent strong laser pulse then maps this stored spin-wave onto a single photon in a cavity mode. 
The collective atom-cavity interaction enables high internal retrieval efficiencies up to $\chi=0.75 \pm 0.02$ while preserving quantum correlations, a threefold improvement over free-space configuration. While strong coupling between atoms and the cavity is typically observed through the transmission or reflection spectrum of the system, we demonstrate here that it also manifests as vacuum Rabi splitting in the retrieval efficiency spectrum. 
The measurement of $\chi$ across different atoms-cavity detunings reveals the typical avoided crossing of a strongly coupled system, in good agreement with our numerical simulations. These findings provide new insights into cavity-enhanced spin-wave readout in cold atomic quantum memories, offering a detailed study of collective spin excitations strongly coupled to a cavity.

\section{Cavity-coupled atomic ensemble}
When a collection of atoms is coupled to an optical cavity mode, the interaction between the atomic ensemble and the quantized electromagnetic field in the cavity mode changes the absorption properties of the system.
Throughout this paper, we consider an ensemble of $N$ atoms coupled to a cavity with resonance angular frequency $\omega_{\mathrm{c}}$, tuned close to the atomic transition ${\ket{g} \xrightarrow{} \ket{e}}$ with angular frequency $\omega_{\mathrm{a}}$, at detuning ${\Delta_c = \omega_{\mathrm{a}} - \omega_{\mathrm{c}}}$.
Neglecting ground-state atom-atom interactions, and using the rotating wave approximation (weak probe drive), the absorption spectrum of our system can be accurately described by the Tavis-Cummings Hamiltonian. In the cavity rotating frame, this is given by: 
\begin{equation}
    \label{eq:hamiltonian}
    H/\hbar = \Delta_c \hat{\sigma}_{ee} - g(\hat{\sigma}_{ge}^{\dagger} a + \hat{\sigma}_{ge} a^{\dagger})\text{.}
\end{equation}
This expression corresponds to that of the single-atom Jaynes-Cummings Hamiltonian, except that the single-atom operators are now replaced by the collective operators $\hat{\sigma}_{ge}^\dagger= \sum_{j=1}^{N} \ket{e_j}\bra{g_j}$, and $\hat{\sigma}_{ee}= \sum_{j=1}^{N} \ket{e_j}\bra{e_j}$.
The electromagnetic field in the optical mode of the cavity, described by the field raising and lowering operators $a^{\dagger}$ and $a$, is coupled to the atomic ensemble with coupling strength $g=g_{0}\sqrt{N}$, where $g_{0}$ is the single atom coupling strength.
This Hamiltonian conserves the number of excitations in the system by exchanging the number of photons in the cavity and the number of collective excitations.
For a single excitation in the system, there are two eigenstates $\ket{\pm}$ with angular frequency 
\begin{equation}
    \label{eq:eigenfreq}
    \omega_{\pm} - \omega_{\mathrm{a}} = \frac{\Delta_c}{2} \pm \sqrt{ g^{2} + \left(\frac{\Delta_c}{2}\right)^{2}}\text{.}
\end{equation}
When $\Delta_c = 0$, the interaction lifts the degeneracy and the levels are split by $2g$, a phenomenon known as the collective vacuum-Rabi splitting~\cite{Zhu1990, Agarwal1984}.

Our experimental setup is based on a laser-cooled ensemble of Rubidium-87 atoms coupled to a single mode of a low-finesse cavity ($\mathcal{F}=23$). 
The optical resonator is placed outside the vacuum system, and consists of three planar mirrors placed in a triangular configuration as illustrated in \autoref{fig:setup} (a).
Our cavity has length ${l = \SI{88}{\centi\meter}}$, free spectral range of ${\Delta\nu=\SI{340}{\mega\hertz}}$, TEM$_{00}$ waist $w_{0} = \SI{69}{\micro\meter}$ and decay rate ${\kappa = 2\pi\cdot\SI{7.25}{\mega\hertz}}$.
In-coupling or out-coupling is done via one partially reflective mirror ($86\%$ reflection), leading to a total escape efficiency of ${\eta_{\mathrm{esc}}=56\%}$ mainly limited by the optical losses of the glass vacuum cell ($11\%$ round trip loss, with $8\%$ from the cell only).
The position of one of the highly reflective mirrors can be slightly adjusted by changing the voltage of a piezo-actuator, allowing us to tune the resonance frequency of the optical resonator.
Using an ensemble of atoms in the cavity mode allows us to reach the collective strong coupling regime defined by ${g > \Gamma, \kappa}$, where ${\Gamma =  2\pi\cdot\SI{6.07}{\mega\hertz}}$ is the natural linewidth of the considered atomic transition and $g$ is the coupling strength between the atoms and the cavity.

To measure the coupling strength $g$ of our system, we first set the cavity resonance to the ${\ket{5S_{1/2}, F=2} \xrightarrow{} \ket{5P_{3/2}, F=2}}$ transition ($\Delta_c = 0$).
We then send a probe through the out-coupling/in-coupling mirror and measure its reflection off the cavity. 
When we scan the probe detuning around this transition, as depicted in \autoref{fig:setup} (c), we observe a splitting of the empty cavity resonance, indicative of the collective vacuum-Rabi splitting.
The splitting is used to measure $g$ and we obtain ${g = g_{0}\sqrt{N} \sim 2\pi \cdot \SI{15}{\mega\hertz}}$.
For this measurement, Zeeman pumping drives our atomic population to the state $\ket{5S_{1/2},F=2,m_{F}=+2}$, while the probe is set to $\sigma^{-}$ polarization, enabling us to drive a precisely defined transition with a well defined dipole matrix element. 
This is used to compute the single-atom coupling strength $g_{0} = 2\pi\cdot\SI{23}{kHz}$, which provides an estimate for ${N \sim 4\times10^5}$, the number of atoms in our cavity mode.
Our data is well reproduced by solving the steady state optical Bloch equations, represented by the solid lines in \autoref{fig:setup}(c).

\section{Cavity enhancement of spin-wave read-out in the quantum regime}
We then use our cavity-coupled atomic ensemble as a source of heralded single photons using the Duan-Lukin-Cirac-Zoller (DLCZ) protocol \cite{Duan2001}, as represented in \autoref{fig:setup}(b). 
The relevant atomic levels are ${\ket{g} = \ket{5S_{1/2}, F=2}}$, ${\ket{e} = \ket{5P_{3/2}, F=2}}$ and  ${\ket{s} = \ket{5S_{1/2}, F=1}}$.
We start by sending a train of write pulses towards the atoms, detuned by ${\Delta_w / (2 \pi) =  \SI{-40}{\mega\hertz}}$ from the $\ket{g} \xrightarrow{} \ket{e}$ transition. 
With low probability, a photon (called the write-out photon) can be generated due to Raman scattering, with detuning $\Delta_w$ from the $\ket{s} \xrightarrow{} \ket{e}$ transition. 
A polarized beam splitter (PBS) is placed inside the cavity, such that the write-out photon directly reflects off the PBS and does not see the cavity.
These write-out photons are sent to a Fabry-Perot cavity to filter-out unwanted frequencies and are subsequently detected on a single photon counting module with probability $p_w$.
This probability can be varied by changing the power of the write laser pulses.
In the ideal case, a detection of this photon heralds a collective spin excitation $\ket{\Bar{s}}$ in our atomic cloud. 
A collective spin excitation, or spin-wave, is a single spin excitation shared by all the atoms:
\begin{equation}
     \ket{\Bar{s}} = \frac{1}{\sqrt{N}}\sum_{j=1}^{N} e^{i\phi_{j}} \ket{g_{1}...s_{j}...g_{N}}\text{,}
\end{equation}
where $\phi_{j}$ is a phase factor for the j$^{\text{th}}$ atom.
This collective state is usually long-lived (half-time $\tau = \SI{15}{\micro\second}$ in our case) and can be efficiently mapped to a single photon called the read-out photon, whose emission directionality is governed by phase-matching conditions.
In our configuration, this corresponds to an emission in the direction opposite to that of the write-out photon.
This mapping is done by sending a read pulse -- typically $\SI{500}{\nano\second}$ after the write pulse -- with detuning $\Delta_{r}$ relative to the $\ket{s} \xrightarrow{} \ket{e}$ transition, producing the emission of a read-out photon in our cavity mode.
We collect this photon in a single mode fiber ($75\%$ coupling efficiency) and send it to a single photon detection platform.
The use of an optical resonator close to resonance with the read-out photon enhances the conversion from the collective excitation to the single photon, as demonstrated in seminal experiments \cite{Simon2007a, Bimbard2014a, Bao2012}.

\begin{figure}[t]
    \centering
    \includegraphics[width=0.48\textwidth]{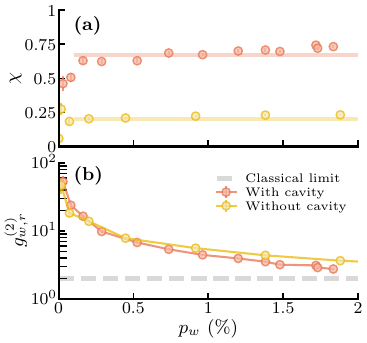}  
    \caption{Cavity-enhanced photon retrieval from spin waves (a): Conversion efficiency $\chi$ of a single collective excitation into a single photon in the cavity mode, as a function of write-out detection probability $p_w$, with (red) and without (yellow) cavity. Solid lines are a guide to the eye at $\chi=0.67$ and $\chi=0.2$ We observe the strong cavity enhancement leading to an increase in $\chi$ when we use the cavity. For both figures, error bars are calculated from Poissonian statistics and are generally smaller than the data points. (b): Cross correlation $g^{(2)}_{w,r}$ between the write-out and read-out photon, as a function of write-out detection probability $p_w$, with (red) and without (yellow) cavity. All our measurements are obtained above the classical limit, given by the Cauchy-Schwarz inequality (see text).
    }
    \label{fig:fig2}
\end{figure}

The conversion efficiency $\chi$ of a single collective excitation into a photon in the cavity mode can be estimated by measuring statistics of the photon' detections. 
We first calculate the conditional retrieval efficiency ${p_{r|w}^{c}}$ of our apparatus by correcting the read-out photon detection probability ${p_{r|w}}$ (conditionned on a prior write-out detection) taking into account the background noise probability $p_{\mathrm{b}}$ and the total transmission on our optical path ${\eta_{\mathrm{tot}}=\eta_{\mathrm{esc}}\eta_{\mathrm{t}}\eta_{\mathrm{d}}}$ following $ {p_{r|w}^{c}=p_{r|w}/\left[ \eta_{\mathrm{tot}}\left( 1 - p_{\mathrm{b}}/p_{w} \right) \right] }$ \cite{Bao2012}. 
Here, $\eta_{\mathrm{esc}}=0.56$ is the escape efficiency of our cavity, $\eta_{\mathrm{t}} = 0.53$ is the total transmission from the cavity to the detector and $\eta_{\mathrm{d}} = 0.45$ is our detector efficiency.
To calculate the conversion efficiency $\chi$, we first need to measure the cross-correlation between the write-out and read-out photons $g^{(2)}_{w,r}=p_{w,r}/p_{w}p_{r}$, where $p_{w}(p_{r})$ is the raw detection probability of the write (read) detector, and $p_{w,r}$ is the unheralded coincidence probability.
This quantity indicates the ratio between coincidence events due to a photon pair detection and random events due to accidentals.
Therefore, we obtain the intrinsic conversion efficiency of a single collective excitation into a single photon in our cavity $\chi$ by subtracting accidental coincidences from the retrieval efficiency of our system ${\chi = p_{r|w}^{c}(1 - 1/g^{(2)}_{w,r})}$ \cite{Bao2012}.
We plot this quantity as a function of write-out detection probability $p_{w}$ in \autoref{fig:fig2}(a), and compare it to the case in which we perform this experiment without a cavity.
As expected, we observe a strong enhancement of the conversion efficiency when using the cavity, reaching values up to $\chi=0.75 \pm 0.02$ compared to $\chi=0.20 \pm 0.02$ without using it.
Similar intra-cavity efficiencies were obtained in \cite{Simon2007a,Bao2012}.
We emphasize here that the read-out photon might be lost in the cavity due to internal losses, and the overall probability that we get a read-out photon out of the cavity from a single spin wave is thus $\chi .\eta_{\mathrm{esc}} = 0.42 \pm 0.01$. By considering the additional optical losses due to fiber in-coupling (82\% coupling) and a chopper-based locking system (64\% transmission), the total in-fiber generation efficiency in our current setup is not really improved compared to the free-space case.

We also plot the corresponding values of $g^{(2)}_{w,r}$ in each case, showing that all these measurements were obtained in the quantum regime by at least one standard deviation.
Here, using the cavity does not significantly improve the cross correlation values because the spurious read-out noise is also enhanced by the cavity. 
The classical limit, represented by the grey horizontal line in \autoref{fig:fig2}(b), corresponds to the upper bound of the Cauchy-Schwarz inequality given by ${g^{(2)}_{w,r} < \sqrt{g^{(2)}_{w,w}g^{(2)}_{r,r}}}$. 
In the ideal case, the autocorrelation functions $g^{(2)}_{w,w}$ and $g^{(2)}_{r,r}$ correspond to those of a thermal state and are both equal to 2, leading to the upper bound of $g^{(2)}_{w,r}<2$.
However, experimental imperfections such as noise or power fluctations can sometimes alter the value of the autocorrelation functions, leading to a different value for the classical limit.
We experimentally measured the values of $g^{(2)}_{w,w}$ and $g^{(2)}_{r,r}$ (see \autoref{sec:appendixC}), leading to a new classical bound of $g^{(2)}_{w,r}<1.52$.

For the measurements presented in \autoref{fig:fig2}, the cavity was locked at ${\Delta_c / (2\pi) = \SI{-14}{\mega\hertz}}$, similar to the read pulse detuning ${\Delta_r = \Delta_c}$.
The write pulse is a $\SI{50}{\nano\second}$-wide gaussian waveform with a tunable peak power up to $\SI{2.4}{\milli\watt}$, corresponding to a write-out photon detection probability ${p_w}$ between 0 and $\SI{2}{\percent}$.
The read pulse is a $\SI{250}{\nano\second}$-wide gaussian waveform with $\SI{150}{\micro\watt}$ peak power.

\section{Conversion efficiency spectrum}
\begin{figure*}[t]
    \centering
    \includegraphics[width=0.98\textwidth]{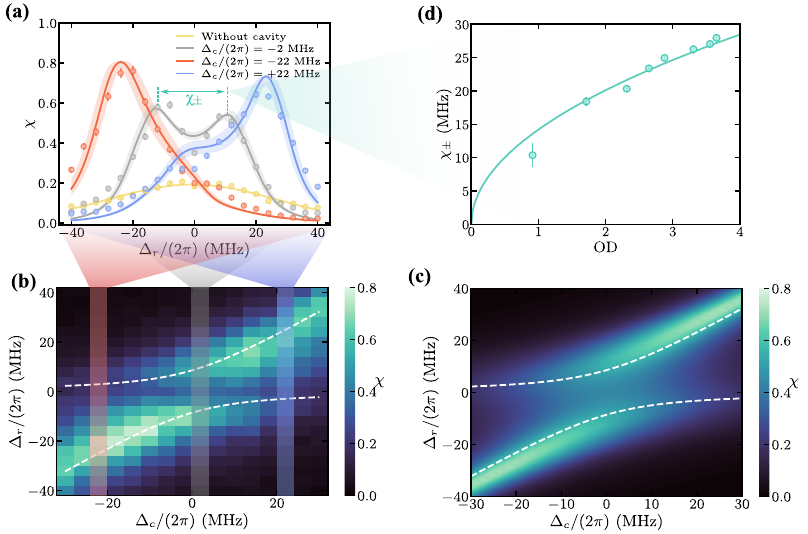}  
    \caption{Observation of vacuum-Rabi splitting of spin wave to single photon conversion. (a): Conversion efficiency $\chi$ of a single collective excitation into a single photon in the cavity mode, as a function of read pulse detuning $\Delta_{r}$, for different cavity detuning. When the cavity is close to atomic resonance (grey), we observe two peaks of maximal efficiency separated by $\chi_\pm$, resulting from the collective atom-cavity interaction, while for larger $\Delta_c$, the atom-cavity system is decoupled and the $\chi$ is maximum for $\Delta_r \approx \Delta_c$. The conversion efficiency $\chi$ is much smaller when the ensemble is not placed inside a cavity (yellow). (b-c): Measurement (b) and simulation (c) of $\chi$  as a function of $\Delta_c$ and $\Delta_{r}$. The maxima closely follows the eigenfrequencies of the three-level Hamiltonian (white dashed line) given in \autoref{eq:eigenfreq3level} which result from the collective atom-cavity coupling. (d): Frequency separation of the conversion efficiency peaks $\chi_\pm$ for $\Delta_c=0$ as a function of the single-pass optical depth (OD). Here the solid line is a fit with a function proportional to $\sqrt{OD}$.
    }
    \label{fig:fig3}
\end{figure*}
We then measured the spectrum of the conversion efficiency $\chi$ by scanning the cavity  $\Delta_c$ and the read pulse $\Delta_r$ detuning, for a given value of $p_\text{w}\approx 1\%$ (see  \autoref{fig:fig3}).
When the cavity is resonant with the atomic transition $\ket{g} \xrightarrow{} \ket{e}$, we observe two distinct peaks of enhanced conversion efficiency (grey circles), a consequence of the collective vacuum-Rabi splitting.
Vacuum-Rabi splitting is generally described in the context of transmission or emission of light in an atom/cavity system, as reported for example in the measurement of \autoref{fig:setup} (c).
Here, we indirectly access this splitting by measuring $\chi$, which is linked to the absorption of light in the system through the optical depth around the $\ket{g} \xrightarrow{} \ket{e}$ transition.
In a first approximation, the retrieval efficiency increases when there is more absorption in the system due to improved collective enhancement \cite{Gorshkov2007}.
Therefore, we expect the optimum retrieval efficiency to coincide with the maxima in the absorption spectrum of the system.
This discussion only holds for small values of optical depth since -- for large values of OD, above 100 typically -- off-resonant scattering from neighboring states becomes important and leads to a reduction of efficiency \cite{Vernaz-Gris2018}.

To simulate the spectrum of $\chi$, we use a three-level extension of the Hamiltonian in Eq. \eqref{eq:hamiltonian} and numerically solve the time-dependent optical Bloch equations in the low excitation limit (see \autoref{simulations} and \cite{Gorshkov2007a}).  Using the rotating wave approximation, in an effective rotating frame the Hamiltonian reads \cite{Gorshkov2007a}:

\begin{equation}
\label{eq:Hamiltonian3level}
\begin{split}
    H/\hbar = & \Delta_c \hat{\sigma}_{ee} + \delta \hat{\sigma}_{ss} - (g\hat{\sigma}_{ge}^{\dagger} a + \Omega(t)\hat{\sigma}_{se}^{\dagger} + \mathrm{h.c}) \text{.}
\end{split}
\end{equation}
Here $\Omega(t)$ is the slowly-varying Rabi frequency of the Read pulse that couples the $\ket{s}$ and $\ket{e}$ states, while $\delta=\Delta_c - \Delta_r$ is the two-photon detuning and $\hat{\sigma}_{se}^\dagger= \sum_{j=1}^{N} \ket{e_j}\bra{s_j}$, and $\hat{\sigma}_{ss}= \sum_{j=1}^{N} \ket{s_j}\bra{s_j}$. The solid lines in \autoref{fig:fig3}(a) are fit to the data using the simulated spectrum with $\Omega_0$, $g$, and $\Delta_c$ as free fitting parameters, where $\Omega_0$ is the amplitude of the Gaussian temporal profile of the read pulse. The fitted parameters $\Delta_c$, $\Omega_0$, and $g$ for the three curves shown in \autoref{fig:fig3}(a) are $\Delta_c / (2 \pi) = \{-1.5(2), 17.6(4),-24.2(3)\}\, \mathrm{MHz}$, $\Omega_0 / (2 \pi) = \{4.8(1), 6.2(4),10.9(5)\}\, \mathrm{MHz}$ and $g / (2 \pi) = \{15.8(1), 15.4(3), 11.4(2)\}\, \mathrm{MHz}$. The reported errors represent the 95\% confidence interval of the fit, with additional systematic uncertainties of 2 MHz for $\Delta_c$ due to cavity stabilization fluctuations and 3\% for $\Omega_0$ from read pulse power variations. The model provides a good representation of the data, however, we expected to find consistent values of $\Omega_0$ and $g$ across all three curves. The observed variations likely stem from limitations in our model, which does not account for the full complex level structure of $^{87}$Rb or various experimental imperfections such as magnetic field fluctuations,  polarization misalignment and inhomogeneous initial population in the Zeeman sub-levels. 

The peaks observed in the conversion efficiency closely follow the steady-state eigenvalues of the Hamiltonian: 
\begin{equation}
\label{eq:eigenfreq3level}
    \Delta_{\pm} = \frac{\delta}{2} \pm \frac{1}{2} \sqrt{g^2 + \Omega^2 + \delta^2}
\end{equation}

For large $\Delta_c$, the system is decoupled and $\Delta_{\pm}$ approaches the bare atom and cavity resonances, while for $\Delta_c \approx 0$, $\Delta_{\pm}$ are the eigenfrequency of the atom-cavity dressed states, separated by $2\sqrt{\Omega^2+g^2}$. We confirm this by measuring and simulating the full spectrum for different cavity detunings $\Delta_c$. Both data and simulations, reported in \autoref{fig:fig3}(b-c), show that the position of the maxima overlaps with the eigenvalues of Eq. \ref{eq:eigenfreq3level}, shown as a dashed white line. For $\Delta_\pm$ we have used the value obtained for the grey curve in \autoref{fig:fig3}(a) $\{g,\Omega_0\}=2\pi \cdot \{ 15.8(1), 4.8(1)\}\, \mathrm{MHz}$. 
Another interesting result is that optimal conditions are not reached when the cavity is resonant with the atomic transition, but rather achieved when the cavity is slightly detuned.
For large values of detunings $\Delta_{c}$, the efficiency starts to drop slightly due to limitations in read pulse power.

The read-out waveforms resemble the Gaussian read pulse when measured at the maxima of efficiency.
However, when $\Delta_{\text{r}} = 0$, we observe some oscillations in the retrieved waveform, possibly due to oscillatory exchange of excitation between the atoms and the cavity \cite{Bochmann2008}.
Some temporal waveform histograms, alongside a small discussion, are provided in \autoref{sec:appF}.

An interesting observation is that the cavity enhancement of the conversion efficiency $\chi$ differs when the cavity is blue- or red-detuned, as represented in \autoref{fig:fig3}(a) for a detuning $\Delta_c/(2\pi)=\SI{-22}{\mega\hertz}$ (red) and $\Delta_c/(2\pi)=\SI{+22}{\mega\hertz}$ (blue). This could be attributed to the interaction of the cavity and pump field with all the other states of the D2 line of $^{87}$Rb which are not included in the simulation; however, a more detailed model is needed to confirm this. 

The collective atom-cavity field interaction is further characterized by measuring the splitting $\chi_{\pm}$ between the two conversion efficiency maxima at $\Delta_c\approx 0$ as a function of the atomic cloud's single-pass optical depth (OD). Our data are reported in \autoref{fig:fig3}(d) where we scan the OD by adjusting the trapping beam power during magneto-optical trap loading. As the OD scales linearly with atom number $N$, we observe a trend proportional to $\sqrt{OD}$ which confirms the observation of vacuum-Rabi splitting in the read-out photon emitted by the atomic spin-wave.

\section{Conclusion}

In conclusion, our study presents a comprehensive analysis of cavity-enhanced mapping of collective excitations into single photons, revealing a notable conversion efficiency of up to $\chi=0.75 \pm 0.02$. 
We have compared this against the free-space scenario, shedding light on the advantages conferred by our cavity. 
Notably, despite the low finesse of our cavity, working with an atomic ensemble propels us into the strong-coupling regime ($g > \kappa \text{,} \Gamma$), crucial for quantum network applications and efficient light-matter interactions. 
We show that our platform converts spin waves with an improved efficiency compared to the free-space case, over a large range of excitation probability, while being above the classical limit as demonstrated by measuring non-classical correlations between the write-out and read-out photon. 
Furthermore, we demonstrate a correlation between optimal conversion efficiency and the eigenvalues of the system's Hamiltonian. Our results are well reproduced by simulations, enriching our understanding of cavity-enhanced spin wave readout.
Nevertheless, the inherent limitations of our current system, particularly its poor escape efficiency, does not yet allow us to improve the overall detection rates. 
Significant enhancements are necessary to increase the in-fiber read-out efficiency and realize a practical truly efficient light-matter entanglement source. 
Strategies to mitigate losses include placing the cavity in vacuum to minimize internal losses, albeit at the expense of increased dephasing due to increased beam angle between the write/read beams and the photon paths.
Alternatively, advancements in highly transmissive optics, enabled by specialized coatings reducing Rubidium deposition on chamber walls could offer a promising avenue for future improvements.

\section{Acknowlegements}
 This project received funding from the European Union Horizon 2020 research and innovation program under Grant Agreement No. 899275 (DAALI), from the Government of Spain (PID2019-106850RB-I00;  CEX2019-000910-S [MICIN/ AEI/10.13039/501100011033]; BES-2017-082464), from the MICIN with funding from European Union NextGenerationEU PRTR (PRTR-C17.I1), from the Gordon and Betty Moore Foundation through Grant No. GBMF7446 to H. d. R, from Fundaci\'o Cellex, Fundaci\'o Mir-Puig, and from Generalitat de Catalunya (CERCA, AGAUR). ED acknowledges support from the ``laCaixa'' Foundation through the fellowship ID100010434, code LCF/BQ/PI23/11970036.

\section{Disclosure}
The authors declare no conflict of interest.

\appendix
\section{Cavity properties}
All the cavity properties are defined by the mode waist, length and losses. 
In the following, we list all the properties of our optical cavity.
The intra-cavity losses are $L = 11\%$, mainly due to the rubidium deposition on the vacuum chamber glass.
The reflectivity of the in/out-coupling mirror is chosen to be $R=86\%$, and the cavity length is $l=\SI{88}{\centi\meter}$. 
The escape efficiency, or the probability that a photon traveling inside the cavity escapes it through the out-coupling mirror, is given by ${\eta_{\mathrm{esc}}= (1-R)/(1-R+L) = 56\%}$. 
The cavity finesse is ${\mathcal{F}=\pi \frac{ \left[ R (1 - L)  \right]^{1/4} }{ \left( 1 - \left[R (1 - L)\right]^{1/2}\right)} = 23.5}$, the free spectral range is defined by $\Delta_{\nu} = c/l = \SI{340}{\mega\hertz}$ and the cavity decay rate is ${\kappa = \pi \Delta_{\nu}/\mathcal{F} = 2\pi\cdot\SI{7.25}{\mega\hertz}}$.
The cavity mode waist is $w_{0} = \SI{69}{\micro\meter}$.

When the cavity is coupled to an atomic transition with wavelength $\lambda$, the strength of this coupling for a single atom is given by
\begin{equation}
    g_{0} = \abs{d} \sqrt{\frac{2c}{\hbar \epsilon_{0} \lambda l w_{0}^{2}}}\text{,}
\end{equation}
where $\epsilon_{0}$ is the vacuum permittivity and $d$ is the dipole matrix element $\bra{g}e\mathbf{r}\ket{e}$.

\section{Cavity reflection spectrum}
In the following, we explain how we derived the reflection spectrum of our system that is used to fit our data from \autoref{fig:setup} (c).
We followed the derivation presented in the appendix of \cite{Vaneecloo2022a}.
Knowing the many-atom Hamiltonian of the system written in \autoref{eq:hamiltonian}, we can write the master equation using the two decay channels of the system, where $\gamma = \Gamma / 2$, and $\kappa$ is the cavity field decay rate.
\begin{equation}
\begin{split}
    \frac{\mathrm{d}\rho}{\mathrm{d}t} = & -\frac{i}{\hbar} \left[ H , \rho \right] + 2 \kappa a \rho a^{\dagger} - \frac{1}{2} \{ 2 \kappa a a^{\dagger}, \rho \} \\
    & + 2 \gamma \Bar{e} \rho \Bar{e}^{\dagger} - \frac{1}{2} \{ 2 \gamma \Bar{e} \Bar{e}^{\dagger}, \rho \} 
\end{split}
\end{equation}
We derive and solve for the steady-state case, for the resonant cavity ($\omega_c = \omega_a$). 
From the steady state Bloch equations, one can obtain the susceptibility of the medium, leading to the reflectivity:

\begin{equation}
\label{eq:R}
    \mathcal{R}(\omega) = \abs{
    1-\frac{i2\kappa_{0}}{\omega - \omega_{a} + i \kappa - g^2 / (\omega - \omega_a + i \gamma)}
    }^2
    \text{,}
\end{equation}
where $\kappa_{0}$ is the decay rate of the cavity through the in/out-coupling mirror. For an empty cavity, $g=0$ and the reflection is minimized if $\kappa / \kappa_0 = 2 $, a condition known as impedance matching. 
In our case, $\kappa / \kappa_0 = 1.91$.
When the atomic ensemble is loaded inside the cavity, we use \autoref{eq:R} to fit the data of \autoref{fig:setup} (c) where the only free parameter is the number of atoms $N$ in $g = g_0 \sqrt{N}$.

\section{Cauchy-Schwarz inequality}
\label{sec:appendixC}
\begin{figure}[t]
    \centering
    \includegraphics[width=0.40\textwidth]{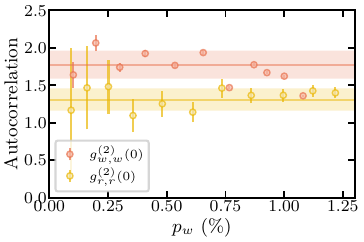}  
    \caption{Unheralded autocorrelation measurements for the write-out field (red), and read-out field (yellow). Solid lines and shaded area indicate the weighted average and standard deviation. The product of these two values is below the theoretical value of 4 (see text). Error bars are derived from Poissonian statistics.}
    \label{fig:appendix1}
\end{figure}
The non-classical character of the correlation between the write-out and the read-out photon is verified by observing a violation of the Cauchy-Schwarz inequality for coincidences of the involved fields.
The inequality states that the cross-correlation $g^{(2)}_{w,r}$ of two classical fields must also manifest
itself in the autocorrelations $g^{(2)}_{w,w}$ and $g^{(2)}_{r,r}$:
\begin{equation}
    {\frac{\left(g^{(2)}_{w,r}\right)^2}{g^{(2)}_{w,w}g^{(2)}_{r,r}}<1}\text{.}
\end{equation}
The two light fields produced by the DLCZ protocol are expected to be in a thermal state \cite{Duan2001} $g^{(2)}_{r,r}= g^{(2)}_{w,w} = 2$ leading to a classical limit of $g^{(2)}_{w,r}<2$.
However, experimental imperfections can sometimes alter the values of the unheralded autocorrelations, thereby changing the value of the classical limit.
For this reason, we measured these unheralded autocorrelations for different excitation probabilities $p_w$.
As represented in \autoref{fig:appendix1}, close to all the autocorrelations values are below 2 in this range.
By taking the weighted average of these values, non-classical correlations are obtained in the case where $g^{(2)}_{w,r}>1.52$.
Thus, the theoretical bound of $g^{(2)}_{w,r}=2$ marking the transition from classical to non-classical correlations is used as a conservative bound, and is represented by the gray dashed line in \autoref{fig:fig2} (a).

\section{Polarization scheme}
\label{sec:appendixD}
\begin{figure}[t]
    \begin{minipage}[c]{0.8\linewidth}
   \def\svgwidth{\columnwidth}
   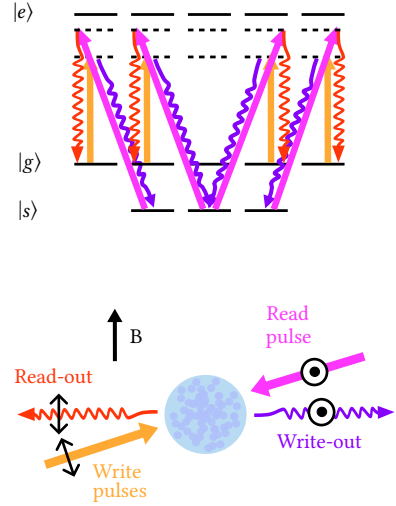
    \end{minipage}
    \caption{The polarization scheme. We apply a bias field perpendicular to the photon mode. The write-out is emitted in both $\sigma^{+}$ and $\sigma^{-}$ channels, corresponding to a vertical polarization in the optical table plane, and thus decouple from the cavity via a polarizing beam splitter. The read-out mode is emitted in a $\pi$ configuration, corresponding to a horizontal polarization in the table plane, and is thus coupled to the cavity.}
    \label{fig:fig6}
\end{figure}

For the DLCZ experiment, the relevant atomic levels are ${\ket{g} = \ket{5S_{1/2}, F=2}}$, ${\ket{e} = \ket{5P_{3/2}, F=2}}$ and  ${\ket{s} = \ket{5S_{1/2}, F=1}}$. Due to the geometry of the cavity, write-out photons and read-out photon must have opposite linear polarization. For this reason, no Zeeman pumping is applied and a small bias field is applied at the cloud, perpendicular to the photon propagation (see \autoref{fig:fig6}). After excitation with horizontally polarized write pulse light, only vertically polarized write photons are selected by polarization filtering. Those photons result from a superposition of $\sigma^{+}$ and $\sigma^{-}$ polarized decay channels. They are not resonant to the cavity and are reflected on the cavity PBS. After read-out with a vertically polarized read pulse, the read photon is emitted horizontally polarized. This read photon emission is enhanced by the cavity.

\section{Simulations of the conversion efficiency spectrum}
\label{simulations}
To simulate the spectrum of the conversion efficiency, we follow the derivation described in \cite{Gorshkov2007a}, and extend it to a case where a non-zero two photons detuning $\delta$ is considered. Using the Hamiltonian in \autoref{eq:Hamiltonian3level}, we derive the equation of motion for the system operators. As we work in the low excitation regime, we assume   $ \hat{\sigma}_{ss}\approx\hat{\sigma}_{ee}\approx\hat{\sigma}_{es}\approx 0$, and $\hat{\sigma}_{gg}= \sum_{j=1}^{N} \ket{g_j}\bra{g_j}\approx N$. we can then define the polarization $\hat{P} = \hat{\sigma}_{ge}/\sqrt{N}$ and spin-wave $\hat{S} = \hat{\sigma}_{gs}/\sqrt{N}$ operators and find the following coupled differential equations:

\begin{align}
\dot{\hat{a}} &= -\kappa \hat{a} + i\sqrt{N}g_0 \hat{P} \\
\dot{\hat{P}} &= -(\gamma + i\Delta)\hat{P} + i\Omega(t) \hat{S} + i\sqrt{N}g_0 a \\
\dot{\hat{S}} &= -(\gamma_s + i\delta)\hat{S} + i\Omega(t) \hat{P}
\end{align}

Here $2 \kappa = 2 \pi \cdot \SI{7.25}{\mega \hertz}$ is the cavity decay rate, $\gamma = 2 \pi \cdot \SI{3.03}{\mega \hertz}$ is the atomic coherence decay rate, and $\gamma_s = \SI{6.7}{\kilo \hertz }$ is the spin-wave decoherence rate. As we are interested in the conversion efficiency, we assume $\hat{S}(0)=1$ while $\hat{a}(0) = \hat{P}(0) = 0$ as initial boundary conditions and proceed to numerically solve the coupled differential equations. The intrinsic conversion efficiency is then given by $\chi = 2\kappa \int_0 ^\infty |\hat{a}(t)| ^2$. 

\section{Single-photon waveforms}
\label{sec:appF}
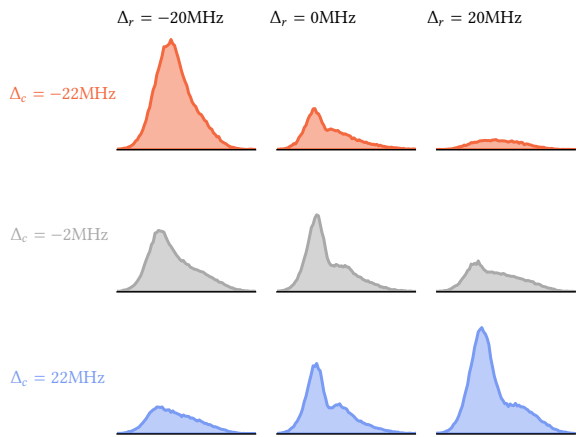
\begin{figure}[t]
    \begin{minipage}[c]{0.9\linewidth}
   \def\svgwidth{\columnwidth}
\begingroup%
  \makeatletter%
  \providecommand\color[2][]{%
    \errmessage{(Inkscape) Color is used for the text in Inkscape, but the package 'color.sty' is not loaded}%
    \renewcommand\color[2][]{}%
  }%
  \providecommand\transparent[1]{%
    \errmessage{(Inkscape) Transparency is used (non-zero) for the text in Inkscape, but the package 'transparent.sty' is not loaded}%
    \renewcommand\transparent[1]{}%
  }%
  \providecommand\rotatebox[2]{#2}%
  \newcommand*\fsize{\dimexpr\f@size pt\relax}%
  \newcommand*\lineheight[1]{\fontsize{\fsize}{#1\fsize}\selectfont}%
  \ifx\svgwidth\undefined%
    \setlength{\unitlength}{195.41544036bp}%
    \ifx\svgscale\undefined%
      \relax%
    \else%
      \setlength{\unitlength}{\unitlength * \real{\svgscale}}%
    \fi%
  \else%
    \setlength{\unitlength}{\svgwidth}%
  \fi%
  \global\let\svgwidth\undefined%
  \global\let\svgscale\undefined%
  \makeatother%
  \begin{picture}(1,0.76553096)%
    \lineheight{1}%
    \setlength\tabcolsep{0pt}%
    \put(0,0){\includegraphics[width=\unitlength,page=1]{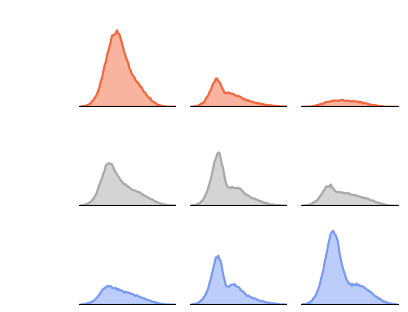}}%
    \put(0.19536137,0.71936088){\color[rgb]{0,0,0}\makebox(0,0)[lt]{\lineheight{1.25}\smash{\begin{tabular}[t]{l}\scriptsize{$\Delta_r = -20 $MHz}\end{tabular}}}}%
    \put(0.45634408,0.71936088){\color[rgb]{0,0,0}\makebox(0,0)[lt]{\lineheight{1.25}\smash{\begin{tabular}[t]{l}\scriptsize{$\Delta_r = 0 $MHz}\end{tabular}}}}%
    \put(0.72500271,0.71936088){\color[rgb]{0,0,0}\makebox(0,0)[lt]{\lineheight{1.25}\smash{\begin{tabular}[t]{l}\scriptsize{$\Delta_r = 20 $MHz}\end{tabular}}}}%
    \put(0.01113839,0.58886952){\color[rgb]{0.95294118,0.40784314,0.24705882}\makebox(0,0)[lt]{\lineheight{1.25}\smash{\begin{tabular}[t]{l}\scriptsize{$\Delta_c = -22 $MHz}\end{tabular}}}}%
    \put(0.01287518,0.34908071){\color[rgb]{0.6627451,0.6627451,0.6627451}\makebox(0,0)[lt]{\lineheight{1.25}\smash{\begin{tabular}[t]{l}\scriptsize{$\Delta_c = -2 $MHz}\end{tabular}}}}%
    \put(0.01287518,0.10344996){\color[rgb]{0.4745098,0.60784314,0.95294118}\makebox(0,0)[lt]{\lineheight{1.25}\smash{\begin{tabular}[t]{l}\scriptsize{$\Delta_c = 22 $MHz}\end{tabular}}}}%
  \end{picture}%
\endgroup%

    \end{minipage} 
    \caption{Temporal mode of the heralded read-out photon, depending on the choice of $\Delta_c$ and $\Delta_r$. All plots here have the same y- and x-scale, representing the number of counts per bin and the time, respectively. The horizontal black line represents $\SI{400}{\nano\second}$. When the read pulse detuning is set to the atomic resonance, we observe some pseudo-oscillations. Our simulations could not reproduce these results.}
    \label{fig:fig7}
\end{figure}

We report on the measured temporal modes of the heralded read-out photons emitted by the DLCZ protocol, as represented in \autoref{fig:fig7}. 
For some configurations of $\Delta_c$ and $\Delta_r$, we observed some oscillations in the read-out photon waveform, while other configuration would lead to a waveform approximately following the Gaussian read pulse shape, especially when measuring at the maxima of efficiency. 

Oscillations in the photon wave packet emitted by a cavity QED system have been previously reported \cite{Bochmann2008} for a single-atom coupled to a high finesse cavity and are well described by the fact that the emission follows the spectrum of the coupled atom-cavity system.
While the emission in free-space is governed by the linewidth of the excited-state, the coupled atom-cavity system evolves with an oscillatory excitation exchange between the atom and the cavity, each of them damped by a different decay constants. 
This leads to an oscillatory behaviour in the retrieved waveform.

Our system, though, is more complex to simulate than that of a single atom coupled to a cavity. First, we have to consider that the read pulse is also interacting with the system during the emission. Collective effects might also play a role, since all the atoms are contributing to the emission.

We observed that the oscillations appeared for almost all values of $\Delta_c$, for a value around $\Delta_r = 0$. We also observe the assymetry that was already visible in \autoref{fig:fig3}(a). We tried to reproduce these waveforms with our model but the results were not conclusive. For now, we do not have a satisfactory answer as to why we observe this asymetry between the red- and blue-detuned case.

\bibliography{DLCZ_in_Cavity.bib}

\end{document}